\documentclass[aoas,MSNbibl,nameyear,dvips]{arximspdf}
\usepackage{upgreek}
\usepackage{graphicx}

\doi{10.1214/11-AOAS470}
\volume{5}
\issue{2B}
\pubyear{2011}
\firstpage{1183}
\lastpage{1206}

\makeatletter
\renewcommand{\epsilon}{\varepsilon}
\DeclareMathAlphabet{\boldmathcaligr}{OMS}{cmsy}{b}{n}

\newcommand{\bTau}{\boldmathcaligr{T}}

\newcommand{\bdelta}{\bolds{\delta}}

\newcommand{\bGamma}{\bolds{\Gamma}}

\newcommand{\bSigma}{\bolds{\Sigma}}

\newcommand{\bmu}{\bolds{\mu}}

\newcommand{\bB}{\mathbf{B}}

\newcommand{\bG}{\mathbf{G}}

\newcommand{\bI}{\mathbf{I}}

\newcommand{\bA}{\mathbf{A}}

\newcommand{\bt}{\mathbf{t}}

\newcommand{\by}{\mathbf{y}}

\newcommand{\mM}{{\mathcal M}}
\newcommand{\eqref}[1]{(\ref{#1})}
\newcommand{\fraca}[2]{{#1}/{#2}}
\makeatother

\begin{document}
\begin{frontmatter}

\title{A nonstationary nonparametric Bayesian approach to dynamically
  modeling effective connectivity in functional magnetic resonance
  imaging experiments}
\runtitle{Dynamic effective connectivity in fMRI}

\begin{aug}
\author[A]{\fnms{Sourabh} \snm{Bhattacharya}\ead[label=e1]{sourabh@isical.ac.in}}
\and
\author[B]{\fnms{Ranjan} \snm{Maitra}\corref{}\thanksref{a2}\ead[label=e2]{maitra@iastate.edu}}
\runauthor{S. Bhattacharya and R. Maitra}
\affiliation{Indian Statistical Institute and Iowa State University}
\address[A]{Bayesian and Interdisciplinary\\ \quad  Research
Unit\\
Indian Statistical Institute\\
203, B. T. Road\\
Kolkata 700108\\
India\\
\printead{e1}} %adresu isvedimo komanda gale!
\address[B]{Department of Statistics\\ \quad  and Statistical Laboratory\\
Iowa State University\\
Ames, Iowa 50011-1210\\
USA\\
\printead{e2}}
\end{aug}
\thankstext{a2}{Supported in part by   NSF CAREER
Grant DMS-04-37555 and by the National Institutes of Health (NIH)
Award DC-0006740.}

% HISTORY:
\received{\smonth{10} \syear{2010}}
\revised{\smonth{2} \syear{2011}}

% ABSTRACT
\begin{abstract}
Effective connectivity analysis provides an understanding of the
functional organization of the brain by studying how activated regions
influence one other. We propose a nonparametric Bayesian approach to
model effective connectivity assuming a dynamic nonstationary
neuronal system. Our approach uses the Dirichlet process to specify
an appropriate (most  plausible according to our prior beliefs)
dynamic model as the ``expectation'' of a set of plausible models
upon which we assign a probability distribution. This addresses
model uncertainty associated with dynamic effective connectivity. We
derive a Gibbs sampling approach to sample from the joint (and
marginal) posterior distributions of the unknowns. Results on
simulation experiments demonstrate our model to be flexible and a
better candidate in many situations. We also used our approach to
analyzing functional Magnetic Resonance Imaging (fMRI) data on a
Stroop task: our analysis provided new insight into the mechanism
by which an individual brain distinguishes and learns about shapes of
objects.\vspace*{2pt}
\end{abstract}

% KEYWORDS

\end{frontmatter}

%s1 ###
\section{Introduction}
\label{secintro}
Functional magnetic resonance imaging (fMRI) is a~non\-invasive
technique for detecting regions in the brain that are activated by
the application of a stimulus or the performance of a task. Although
important neuronal activities are responsible for such activation,
these are very subtle and can not be detected directly. Instead, local changes
during neuronal activity in the flow, volume, oxygen level and other
characteristics of blood, called the blood oxygen level
dependent (BOLD) response, form a~proxy. Much research in fMRI has
focused on identifying regions of cerebral activation in response to
the activity of interest. There is, however, growing interest in
obtaining better understanding of the interactions
between different brain regions during the operation of the BOLD
response. The study of how one neuronal system interacts with another
is called effective connectivity analysis [\citet{Friston94}; \citet{Nyberg01}]. We
illustrate this in the context of obtaining greater insight into how
an individual brain performs a Stroop task, which is also the main
application studied in this paper.\looseness=-1
%s1.1 ###
\subsection{Investigating the attentional control network in a Stroop task}
\label{introACN}
The human brain's  information processing capability is limited, so it
sifts out irrelevant details from task-relevant information
using the cognitive function called \textit{attention}. Specifically,
task-relevant information is filtered out either because of intrinsic
properties of the stimulus (bottom-up selection) or
independently (top-down selection) [\citet{frith01}]. The brain's
preference for task-related information in top-down selection requires
coordination of neural activity via an Attentional Control Network
(ACN) which has systems to process task-relevant and irrelevant
information and also a ``higher-order executive control system'' to
modulate the frequency of neuronal firings in
each [\citet{banichetal00}]. Thus, the higher-order system can execute
top-down selection by increasing neuronal activity in the
task-relevant processing system while
suppressing it in its task-irrelevant counterpart. Many studies have
empirically found the dorsal lateral prefrontal cortex (DLPFC) to be
the main source of attentional control, while the task-relevant and
irrelevant processing sites depend on whether the stimulus is visual,
auditory or in some other form.

\citet{jaensch29} and  \citet{stroop35} discovered that
the brain is quicker at reading named color words (e.g., blue,
yellow, green, etc.) when they are in the concordant color than if they are
in a discordant color. Tasks structured along these lines are now called
Stroop tasks. A much-studied two-phase
experiment [\citeauthor{milhametal02} (\citeyear{milhametal02,milhametal03a});
\citeauthor{Ho03} (\citeyear{Ho03,Ho05});
\citet{milhametal03b};
\citet{Bhattacharya06}] designed around such a task provided the dataset for our
investigation. In the first phase, a subject was trained to
associate each of three unfamiliar shapes with a unique color word
(``Blue,'' ``Yellow'' and ``Green'') with 100\%
accuracy. The second (testing) phase involved alternating six times
between blocks of
eighteen \textit{interference} and eighteen \textit{neutral} trials. The
neutral trial consisted of printing the shape in a neutral
color (white). The interference trial involved presenting the subject
with one of the learned shapes, but printed in a color different from
that learned to be represented by that shape in the learning phase.
The subject's task was to subvocally name the shape's color as
trained in the learning phase, ignoring the color presented in the
testing phase. Each neutral or interference trial consisted of a 0.3-s
fixation cross, a 1.2-s stimulus presentation stage and a 0.5-s
waiting state till the next trial. fMRI images were acquired and
processed to obtain three activated regions, whose averaged
post-processed time series are what we analyze further to investigate
attentional control.
These three regions---also denoted as Regions 1, 2 and 3 in this
paper---were the lingual gyrus (LG), the middle occipital
gyrus (MOG) and the DLPFC, and  were chosen as representatives of
task-irrelevant, task-relevant and executive-control systems,
respectively.
%were the lingual gyrus (LG), the middle occipital gyrus (MOG) and the
%DLPFC and  chosen as representatives of task-irrelevant,
%task-relevant and executive control systems, respectively. We denote
%LG, MOG and DLPFC as Regions 1, 2 and 3 in this paper.
The LG is a visual area for processing color information [\citet{corbettaetal91}],
which in our context is   task-irrelevant [\citet{kelleyetal98}]. The MOG
is another visual area but processes shape information, which is the
task-related information\ (form of the shape) in the experiment.
We refer to \citet{Bhattacharya06} for further details on  data collection
and post-processing, noting here that, as in that and other preceding
papers, the objective is to investigate and to understand the working of
the ACN mechanism in performing a~Stroop task.
%s1.2 ###
\subsection{Background and related work}
%There are two broad approaches to studying effective
%connectivity. These are based upon Structural and Dynamic Causal
%Modelling. Examples of structural causal modelling include structural
%equation modelling  and causal modelling with Bayesian dependency
%graphs (or networks). Dynamic approaches include those formulated in
%continuous time (Dynamic Causal Modelling) and those formulated in
%discrete time  (Dynamic Bayesian Networks and vector autoregression
%models VAR cf, Granger Causality). In general,  structural causal
%modelling ignores dynamics in the observed system. Furthermore, many
%forms of  Dynamic Causal Modelling assume effective connectivity is
%time invariant. There is, however,\ldots}
Structural equation
modeling [\citet{mcintoshandgonzalez94}; \citet{kirketal05}; \citet{pennyetal04b}] and
time-varying parameter regression [\citet{buchelandfriston98}]
are two early approaches that have been used to determine effective
connectivity. In general, both approaches ignore dynamic modeling of
the observed system, even though the latter accounts for temporal
correlation in the analysis. There is, however, strong empirical
evidence [\citet{aertsenandpreibl91}; \citet{Friston94}; \citet{mcintoshandgonzalez94};
 \citet{buchelandfriston98}; \citet{mcintosh00}]
that effective connectivity is dynamic in nature, which means that the
time-invariant model assumed by both approaches  may not be
appropriate.
%Dynamic Causal Modeling (DCM) was therefore proposed by
%vector over time, using stochastic differential equations. It uses
%the ``balloon model'' of \citet{buxtonetal98} along with extensions
%by \citet{stephanetal07} and \citet{fristonetal00}
\citet{Ho05} overcame some of these limitations by modeling
the data using a state-space approach, but did not account for the
time-varying nature of the effective connectivity parameters.

An initial attempt at explicitly incorporating the time-varying nature of
effective connectivity in addition to dynamic modeling of neuronal
systems was by \citet{Bhattacharya06}, who adopted a Bayesian
approach to inference and developed and illustrated their methodology
with specific regard to the ACN mechanism of
the LG, MOG and DLPFC regions in conducting the Stroop task outlined
above. We summarize their model---framing it within the context of
more recent literature in dynamic modeling of effective connectivity---and discuss their findings and some limitations next. In doing so,
we also introduce the setup followed throughout this paper.

%s1.2.1 ###
\subsubsection{Bayesian modeling of dynamic effective connectivity}
\label{subsecnimodel}
Let $y_i(t)$ be the observed fMRI signal (or the measured BOLD response) corresponding to the $i$th
region at time $t$, $i=1,2,\ldots,R$, $t=1,2,\ldots, T$. Specifically,
$y_i(t)$ is some voxel-wise summary (e.g., regional average) of
the corresponding detrended time series in the $i$th region. Following
\citet{Bhattacharya06}, let $x_i(t)$ be the modeled BOLD response [as opposed to the
measured BOLD response, $y_i(t)$], that is,
the stimulus $s(t)$ convolved with
the hemodynamic response function (HRF) $h_i(t)$ for the $i$th region
and time point $t$. In this paper, $h_i(t)$ is assumed to be the
very widely-used standard HRF model of \citet{glover99} which
differences two  gamma functions and has some very appealing properties
vis-a-vis other HRFs [\citeauthor{luetal06} (\citeyear{luetal06,luetal07})]. Then the model for the
observed fMRI signal can be hierarchically specified as
\begin{equation}
y_i(t)=\alpha_i+x_i(t)\beta_i(t)+\epsilon_i(t),
\label{eqhier1}
\end{equation}
where $\alpha_i$ and $\beta_i(t)$ are the baseline
trend and activation coefficients for the $i$th region, the latter at
time $t$. The errors  $\epsilon_i(t)$'s are all assumed to be
independent $N(0,\sigma^2_i)$,
following \citet{Worsleyetal02}. From Bhattacharya, Ho and  Purkayastha [(\citeyear{Bhattacharya06}), page 797], we assume
that $x_i(\cdot)=x(\cdot)$ for $i=1,\ldots,R$, that is, we use the same HRF
$h_i(\cdot)=h(\cdot)$ for each of the $R$~re\-gions. Note that, as argued
in that paper, %As argued in \ctn{Bhattacharya06}, page 797,
this homogeneous assumption on the $x(\cdot)$ is inconsequential
because it is compensated by the $\beta_i(t)$ that are associated with
$x(t)$, and allowed to be inhomogeneous with respect to the different
regions. % via equations  (\ref{eqhier2}) and (\ref{eqhier3}), and
        % compensates for the homogeneity assumption in the HRF.
Also, following \citeauthor{Bhattacharya06} [(\citeyear{Bhattacharya06}), page 799], we
assume that $\sigma^2_i=\sigma^2_{\epsilon}$;
$i=1,\ldots,R$. Actually,~\eqref{eqhier1} is a~generalization of a
very standard model used extensively in the literature---see,
  for example, \citeauthor{lindquist08} [(\citeyear{lindquist08}), equation~(9)] or
\citeauthor{hensonandfriston07} [(\citeyear{hensonandfriston07}), page 179, equation~(14.1)], who use the same
model but with a~constant time-invariant $\beta(t)\equiv
\beta$. (Indeed, as very helpfully pointed out by a reviewer, this
last specification is also the general linear model commonly used to
analyze fMRI data voxel-wise, such as in statistical
parametric mapping and related conventional whole brain activation
studies.) Our specific generalization incorporates time-varying
$\beta(t)$ and follows
\citet{Ho05}, \citet{Bhattacharya06} or
\citeauthor{harrisonetal07} [(\citeyear{harrisonetal07}), cf. page~516, equation~38.18]---note, however, that the
latter model~$\beta(t)$ as a random walk [see equation~38.19,
  page 516, of \citet{harrisonetal07}]. We prefer allowing for time-varying
activation $\beta_i(t)$ in order to address the ``learning'' effect often
reported in fMRI studies whereby strong
activation in the initial stages of the experiment dissipates over
time [\citet{gossletal01};  \citeauthor{milhametal02} (\citeyear{milhametal02,milhametal03a}); \citet{milhametal03b}].
Further modeling specifies the activation coefficient in the $i$th
region at the $t$th time-point in terms of the noise-free BOLD signal
in the other regions at the previous time-point. Thus,
\begin{eqnarray}\label{eqhier2}
  \beta_i(t)=x(t-1) \Biggl[\sum_{\ell=1}^R\gamma_{i\ell}(t)\beta_{\ell}(t-1)
  \Biggr]+\omega_i(t),\\
\eqntext{t=2,\ldots,T;i=1,2,\ldots,R,}
\end{eqnarray}
where $\omega_i(t)$ are independent $N(0,\sigma_\omega^2)$-distributed
errors and $\gamma_{ij}(t)$ is
the influence of the $j$th region on the
$i$th region at time $t$.
Under (\ref{eqhier2}), functionally
specified cerebral areas are not constrained to act independently but
can interact with other regions. % depending on the task or stimulus
                                % involved.
Our objective is to make
inferences on $\gamma_{ij}(t)$ in order to understand the functional
circuitry in the brain as it processes a~certain (in this paper,
Stroop) task.
%Note that in (\ref{eqhier2}) the activation coefficient
%at time $t$, $\beta_i(t)$,
%is modeled conditional on the observed BOLD response at the previous time point,
%$x(t-1)$, effective connective parameters $\gamma_{i\ell}(t)$'s,
%and $\beta_{\ell}(t-1)$'s.
%This implies that changes in the voxel-specific activation coefficients are brought about by
%changes in $x(t-1)$, $\gamma_{i\ell}(t)$'s, and $\beta_{\ell}(t-1)$'s.
%To complete specification of the basic form of our hierarchical model, we still
%need to postulate a model for the effective connectivity parameters.
%Since the latter are not influenced by the observed BOLD signals the
%model should be independent of $x(\cdot)$. Our model for effective connectivity,
%specified in equation (\ref{eqhier3}), and its generalizations thereof,
%throughout respects this (anti-) requirement.

Equations \eqref{eqhier1} and \eqref{eqhier2} together specify one
of many Vector Autoregressive (VAR) models proposed by several
  authors [\citet{harrisonetal03}; \citet{goebeletal03}; \citet{rykhlevskaiaetal06}; \citet{satoetal07};
  \citet{thompsonandsiegle09}; \citet{patriotaetal10}]. To see this, note that for
  $i=1,\ldots,R$, $\beta_i(t-1)$ depends linearly upon $y_i(t-1)$. Hence,
  substituting this in \eqref{eqhier2} yields
  $\beta_i(t)=g_i(y_1(t-1),\break y_2(t-1),\ldots,y_R(t-1))$, for known
  functions $g_i$, which are linear in
\mbox{$y_1(t-1)$}, $y_2(t-1),\ldots,y_R(t-1)$. Then substituting $\beta_i(t)$ in
\eqref{eqhier1}, we see that for each $i=1,\ldots,R$, $y_i(t)$ is a
linear function of  $y_1(t-1),y_2(t-1),\ldots,y_R(t-1)$. Hence, the
vector $\by(t)= (y_1(t),\ldots,y_R(t) )'$ is a linear
function of the vector
$\by(t-1)= (y_1(t-1),\ldots,y_R(t-1) )'$. As a result, our
model is a first order VAR model from the viewpoint of the
responses. It is of first order since~$\by(t)$ depends upon
$\by(t-1)$, given $\by(1),\ldots,\by(t-1)$.  %(just like AR(1)
                                %univariate time-series,
                           %which we  have used in the Dirichlet
                           %process context).
%We show in Section~\ref{methodology} that \eqref{eqhier2} also fits
%the VAR framework.
Moreover, \eqref{eqhier2}~shows that the activation coefficients
$\beta_i(t)$ are modeled as first order VAR; that is,~the
$R$-component vector $ (\beta_1(t),\ldots,\beta_R(t) )'$
depends linearly upon
$ (\beta_1(t-1)$, $\ldots,\beta_R(t-1) )'$.
%and under the marginal process $\bG_0$, the effective connectivity
%parameters $\gamma_{ij}(t)$; $(i,j)=1,\ldots,R$,  also form a VAR
%model of the first order.
% Specifically, letting
%$\bGamma(t)= (\gamma_{ij}(t);i,j=1,\ldots,R )'$, it follows
%that under $\bG_0$, $\bGamma(t)=\rho\bI\bGamma(t-1)+\bdelta(t)$, where
%$\bI$ is the $R\times R$-order identity matrix and
%$\bdelta(t)= (\delta_{ij}(t);i,j=1,\ldots,R )'$. Hence, the
%model for activation coefficients and the (marginal) model for
%effective connectivity parameters are VAR as well. Thus, we note that

VAR models provide an alternative or a substantial
generalization [\citet{friston09}] to the Dynamic Causal Modeling (DCM)
approach proposed by \citet{fristonetal03}, at least in
continuous-time, to model the change of the
neuronal state vector over time, using stochastic differential
equations. In DCM, the observed BOLD signal is modeled as $y_i(t)
= r_i(t) + \beta z_i(t) +\epsilon_i(t)$, where $z_i(t)$
denotes nuisance effects, and
%denotes neural
%activity obtained from the stimulus $s_i(t)$ using a bilinear
%differential (neural state) equation, and
$r_i(t)$ is a modeled BOLD
response obtained by first using a bilinear differential (neural state) equation,
parametrized in terms of effective connectivity parameters and involving $s(t)$,
then subsequently using a ``balloon model''
transformation [\citet{buxtonetal98} or
extensions \citet{fristonetal00}; \citet{stephanetal07}] to the solution of the bilinear differential equation.
DCM thus uses both
$r_i(t)$ as well as the nuisance effects $z_i(t)$ to model the observed BOLD
response, with $r_i(t)$ playing the same role as our $x_i(t)$ with the
exception that the latter is obtained using the more widely-used
\citet{glover99} HRF model. Further, DCM assumes a deterministic
relationship between the different brain regions unlike
\eqref{eqhier2} which allows for noisy dynamics
[\citet{Bhattacharya06}].

\citet{thompsonandsiegle09} contend that VAR models have gained
popularity in recent years because ``the
direction and valence of effective connectivity relationships do not
need to be pre-specified.'' %Such models
                                %relate a temporal dependence of fMRI
                                %activation  As also explained by
                                %them, models ``relate(s) the
                                %time-dependent structure of fMRI
                                %activation of multiple regions
                                %through a multivariate regression of
                                %prior levels of activation on current
                                %levels. Directed connectivity
                                %coefficients between regions are
                                %derived from the off-diagonal
                                %elements of the autoregressive
                                %matrices.''
As such, these models have provided a useful
framework for effective connectivity analysis.% (see, however, the
                                %caution in \citet{friston09}).

\citet{Bhattacharya06} proposed a symmetric random walk model for
$\gamma_{ij}(t)$:
\begin{equation}
\gamma_{ij}(t)=\gamma_{ij}(t-1)+\delta_{ij}(t)   \qquad \mbox{for } i,j=1,2,\ldots,R;t=2,3,\ldots,T,
\label{eqhier3}
\end{equation}
where  $\delta_{ij}$ are independent $N(0,\sigma^2_\delta)$-distributed
errors. In this paper we use $\mM_{\mathrm{RW}}$ to refer to the
model specified by (\ref{eqhier1}), (\ref{eqhier2}) and
(\ref{eqhier3}).
The effective connectivity
parameters $\gamma_{ij}(t)$; $(i,j)=1,\ldots,R$,  also form a VAR
model of the first order. To see this, let
$\bGamma(t)= (\gamma_{ij}(t);i,j=1,\ldots,R )'$. Then  it follows
that $\bGamma(t)=\bI\bGamma(t-1)+\bdelta(t)$, where
$\bI$ is the $R\times R$-order identity matrix and
$\bdelta(t)= (\delta_{ij}(t);i,j=1,\ldots,R )'$, indicating
that $\gamma_{ij}(t)$'s are within the framework of a VAR model.
% The AR(1) modification of (\ref{eqhier3}) proposed in this paper
% is VAR as well, due to similar reasons. In Section
% \ref{secmodellingmethodology} it will be seen that the effective
% connectivity parameters are AR(1), hence, VAR, under the expected
% distribution of our new model based on Dirichlet process. Note
% however, that given a realization of a random distribution from the
% Dirichlet process, such VAR representation does not hold.

\citet{Bhattacharya06} specified
prior distributions on the parameters and hyperparameters of this
model and used Gibbs sampling to learn the posterior distributions of
the unknowns. We refer to that paper for details and for
results on simulation experiments using $\mM_{\mathrm{RW}}$, noting here only
that their Bayesian-derived inference supported ACN theory and, more
importantly, the notion that effective
connectivity is indeed dynamic in the network. Further, they found
that the restricted model with $\gamma_{31}(t) = \gamma_{32}(t) \equiv
0\ \forall  t$ was the best-performer, implying no direct feedback
from the two sites of control (LG and MOG) to the
source (DLPFC). Interestingly, however, and perhaps surprisingly,
their estimated $\gamma_{ij}(t)$'s (see Figure~6 in their
paper) had very little relationship with the nature of the BOLD
response (see Figure~1, bottom panel, in that paper). This is
surprising because from \eqref{eqhier1}, we have
$\beta_i(t)={ (y_i(t)-\alpha_i-\epsilon_i(t) )}/x(t)$, and
similarly for
                    %$\beta_i(t-1)= (y_i(t-1)-\alpha_i-\epsilon_i(t-1) )/x(t-1)$,
                    %substituting this form of
$\beta_i(t-1)$, which when substituted on the right-hand side
of \eqref{eqhier2} makes it independent of $x(\cdot)$. This means
that the effective  connectivity parameters $\gamma_{i\ell}(t)$ depend
upon $\beta_i(t)$, the left-hand side of \eqref{eqhier2}. Since
$\beta_i(t)$ is a function of $x(t)$, it is reasonable to expect
$\gamma_{i\ell}(t)$'s to depend upon $x(t)$, but such a relationship
was not found in \citet{Bhattacharya06}.
% That is why the fact that $\gamma_{ij}(t)$s showing little relationship with $x(t)$ surprised us.
This perplexing finding led us to first investigate robustness of
$\mM_{\mathrm{RW}}$ to even slight misspecifications.

%f1 ###
\begin{figure}

\includegraphics{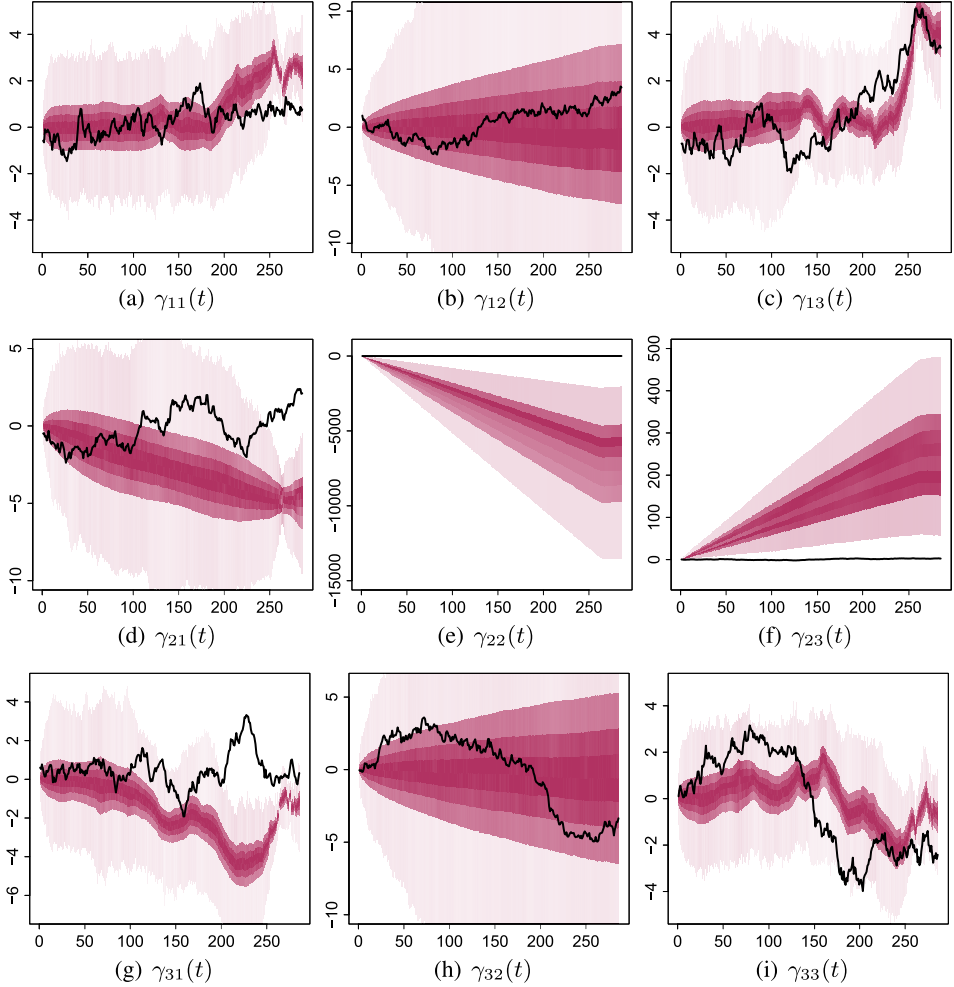}

 \caption{Posterior densities of
 $\gamma_{ij}(t);t=1,\ldots,T;i,j=1,2,3$,
   under model $\mM_{\mathrm{RW}}$
   on data simulated under model $\mM_{\mathrm{RW}'}$.
   The opacity of shading in each region is proportional to the area
   under the density in that region.  The  solid line stands for
   the true values of
 $\gamma_{ij}(t)$.}
 \label{figsimstudyrw}
 \end{figure}

%s1.2.2 ###
\subsubsection{Robustness of the random walk model}
\label{introrobustness}
% However, their modeling of the process by a random walk may be too
% simplistic: indeed,  experiments reported in this paper show that
% estimated effective connectivity parameters degrade considerably
% when the true model is close to nonstationarity, even though it is
% actually  stationary.
%
% We refer to that paper for details and also for results on
% experiments on datasets simulated using $\mM_{\mathrm{RW}}$.  Instead, we
We tested the effect of a~slight departure from $\mM_{\mathrm{RW}}$ by
simulating, instead of from (\ref{eqhier3}), from the following
stationary autoregressive model:
\begin{equation}
 \qquad \gamma_{ij}(t)=0.999\gamma_{ij}(t-1)+\delta_{ij}(t) \qquad  \mbox{for
} i,j=1,2,\ldots,R;t=2,3,\ldots,T.
\label{eqarsim}
\end{equation}
We call this slightly modified model $\mM_{\mathrm{RW}'}$. Here,  $T=285$ and
$R=3$ to match the details of the dataset of Section~\ref{introACN}.
% Note that (\ref{eqarsim}) is almost similar to (\ref{eqhier3}).
% Further, $\gamma_{31}(t) = \gamma_{32}(t) = 0$.
% We refer to the  supplement for complete details on the experiments.
%implying no direct feedback from the two sites of control (LG and
%MOG) to the source (DLPFC).
We fit $\mM_{\mathrm{RW}}$ to data simulated from $\mM_{\mathrm{RW}'}$.
Figure~\ref{figsimstudyrw} displays the estimated posterior
distributions of $\gamma_{ij}(t)$. The marginal \mbox{posterior}
distribution of each $\gamma_{ij}(t)$ is represented here
by eight quantiles, each containing 12.5\% of the distribution:
increased opacity in shading denotes denser regions. Solid lines
represent true values. As seen, many parts of the posterior
distribution have very little coverage of the true effective
connectivity parameters: this finding is also supported by Table
\ref{tableM1} %Table~\ref{tableM1}
which provides the proportion of true values included in the 95\%
highest posterior density (HPD) credible intervals [\citet{Berger85}]
(these are the shortest intervals with posterior probability 0.95).
Thus, performance degrades substantially even though $\mM_{\mathrm{RW}'}$ is
not all that different from $\mM_{\mathrm{RW}}$. Hence, modeling the process by
a random walk may be too restrictive and
thus a better approach may be needed.
%indeed, experiments reported in this paper show that estimated
%effective connectivity parameters degrade considerably when the true
%model is close to nonstationarity, even though it is actually  stationary.
We do so in this paper by embedding an (asymptotically) stationary first
order autoregressive  AR(1) model in a~larger class of models.
% (or more realistically, an asymptotically stationary AR(1) model)   of
%$\{$$\gamma_{ij}(t)$$;t=1,\ldots,T;$ $i,j=1,2,3$$\}$ in a larger
%class of models;
Formally, we employ a Bayesian nonparametric framework using
a Dirichlet Process (DP) prior whose
base distribution is assumed to be that implied by an AR(1)
model. The intuition behind this modeling style is that although one
might expect the actual process to be stationary, the assumption might
be too simplistic, and it is more logical to think of the stationary
model as an ``expected model,'' thus allowing for nonstationarity
(quantified by the DP prior) in the actual model.
%In fact, we demonstrate that if the true model for the $\gamma_{ij}$ is
%stationary AR(1), but close to nonstationarity, then modeling the
%$\gamma_{ij}$'s with either AR(1) model or a random walk model as in
%based on Dirichlet process where the base measure is a stationary
%AR(1) model.
%The rest of our paper is structured as follows.
%
%
% Our model also implies, in contrast with models with independent
% time series as in all previous fMRI-related works, that
% $\{\gamma_{ij}(t);t=1,\ldots,T\}$ are dependent for
% $i,j=1,\ldots,R$, reflected through clusterings, although their
% marginals are stationary AR(1). Simulation studies demonstrated that
% this dependence enhances knowledge regarding the true processes;
% application to the real data revealed interesting and scientifically
% plausible clusterings  among the different effective connectivity
% parameters,  providing important insights into the way different
% brain regions interact.
%
%
Theoretical issues related to the construction\vadjust{\goodbreak} of DP-based
nonstationary processes are discussed in Section \ref{secdpidea}. In
Section \ref{secdynamicdp} we introduce our new modeling ideas
using the developments in Section \ref{secdpidea}.
%Section \ref{secdpgibbs} introduces a Gibbs sampling algorithm for simulating from
%the posterior distributions associated with the new model detailed in Section \ref{secdynamicdp}.
%Possible improvement of the Gibbs sampling algorithm is provided in Section \ref{secconfig}.
The efficacy of the new model is compared with its competitors
%
%t1 ###
\begin{table}
\tabcolsep=0pt
\tablewidth=275pt
\caption{Proportion of true $\gamma_{ij}(t)$ included in the 95\%
  posterior credible intervals obtained using model
  $\mM_{\mathrm{RW}}$ on data simulated using~$\mM_{\mathrm{RW}'}$}
\label{tableM1}
\begin{tabular*}{275pt}{@{\extracolsep{\fill}}lcccccccc@{}}
\hline
$\bolds{\gamma_{11}}$ & $\bolds{\gamma_{12}}$ & $\bolds{\gamma_{13}}$ &
$\bolds{\gamma_{21}}$ & $\bolds{\gamma_{22}}$ & $\bolds{\gamma_{23}}$ & $\bolds{\gamma_{31}}$
& $\bolds{\gamma_{32}}$ & $\bolds{\gamma_{33}}$ \\ \hline
0.99 & 0.99 & 0.91 & 1.0 & 0 & 0.05 & 0.05 & 1.0 & 0.60 \\ \hline
\end{tabular*}
\end{table}
on some simulated datasets in Section~%
\ref{secsimstudy}. The new approach is applied in Section~%
\ref{secreal} to the  dataset introduced in
Section~\ref{introACN} to investigate effective connectivity
between the LG, MOR and DLPFC regions. We conclude in
Section~\ref{secconclusion}
with some discussion.
Additional derivations and further details on
experiments and data analyses are provided in
the supplement [\citet{Bhattacharya11}], whose sections, figures and tables have the prefix
``S-'' when referred to in this paper.\looseness=-1

%s2 ###
\section{Modeling and methodology}
\label{secmodellingmethodology}
\label{secmethodology}
\subsection{A nonstationary Dirichlet process model for time series observations}
\label{secdpidea}
A~random probability measure $\bG$ on the probability space
$(\bGamma,\mathcal B_{\gamma})$ sampled from the Dirichlet Process
(DP) denoted by $\operatorname{DP}(\tau\bG_0)$, and with known distribution $\bG_0$ and
precision parameter $\tau$, can be represented almost surely, using
the constructive method provided in \citet{Sethuraman94}, as
\begin{equation}
\bG\equiv\sum_{k=1}^{\infty}p_k\delta_{\gamma^*_k},
\label{eqsethuraman}
\end{equation}
where $p_1=b_1$ and $p_k=b_k\prod_{\ell=1}^{k-1}(1-b_{\ell}),k=2,3,\ldots,$ with
%$b_k \stackrel{iid}{\sim}Beta(1,\tau)$.
$b_k$'s being independent, identically distributed (henceforth
 i.i.d.) $\upbeta(1,\tau)$
random variables. The values
$\gamma^*_k$ are  i.i.d.  realizations from
$\bG_0$, for $k=1,2,\ldots,$ and are also independent of
$\{b_1,b_2,\ldots\}$. Note that (\ref{eqsethuraman}) implies that
$\bG$ is discrete with probability one, and has expectation $\bG_0$.
%In other words, although the random probability measure $\bG$ is centered around
%$\bG_0$, which may be continuous, it gives probability one to the set of all discrete
%distributions.
DPs thus provide ways to place priors on probability measures.

The dependent Dirichlet process (DDP) is an extension of the DP in the sense
that it allows for a prior distribution to be specified on a set of random
probability measures, rather than on a single
random probability measure. In other words,
the realizations $\gamma^*_k$ can be extended to accommodate an entire time-series domain $\bTau$, such that
$\bGamma^*_{k,\bTau}=\{\gamma^*_{kt};t\in\mathcal\bTau\}$. Following (\ref{eqsethuraman}),
the random process thus constructed  can be represented as
\begin{equation}
\bG^{(\bTau)}\equiv\sum_{k=1}^{\infty}p_k\delta_{\bGamma^*_{k,\bTau}}
\label{eqsethuraman2}
\end{equation}
with form similar to that used for spatial DP models [see
\citet{Gelfand05}]. Note that  $\bGamma^*_{k,\bTau}$ in
\eqref{eqsethuraman2} are realizations of\vspace*{-2pt}
some stochastic process $\bGamma_{\bTau}=\{\gamma_t;t\in\bTau\}$, with
distribution $\bG^{(\bTau)}_0$ for $k=1,2,\ldots.$
Hence, Kolmogorov's consistency holds for $\bGamma_{\bTau}$.
That is, finite dimensional
joint distributions $\{\gamma_t;t\in\bt_T\}$, for ordered time-points $\bt_T=\{t_1,\ldots,t_T\}$,
can be obtained from all finite but higher-dimensional
joint distributions $\{\gamma_t;t\in\bt^*_T\cup\bt_T\}$ (here $\bt^*_T$ is a finite set) specified by the process,
by marginalizing over $\{\gamma_t;t\in\bt^*_T\}$.
Since (\ref{eqsethuraman2}) shows that $\bG^{(\bTau)}$ is specified completely
by the process $\bGamma_{\bTau}$ and $\{p_k;k=1,2,\ldots\}$, and since the latter are independent of $t$, it
follows that Kolmogorov's consistency holds for $\bG^{(\bTau)}$,
providing a
formal setup of
a stochastic process of random distributions.
In particular, for any $t\in\bTau$, $\bG^{(\{t\})}\sim \operatorname{DP}(\tau\bG^{(\{t\})}_0)$
[and admits the representation $\bG^{(\{t\})}\equiv\sum_{k=1}^{\infty}p_k\delta_{\gamma^*_{kt}}$].
The collection of random measures $\bG^{(\bTau)}$\vspace*{1pt} is said to follow the
DDP [see, e.g., \citet{MacEachern00a}; \citet{DeIorio04}; \citet{Gelfand05}].

The process $\bGamma_{\bTau}$ may be a time series that is stationary or---as
adopted in our application and more realistically---asymptotically
so. Indeed, while asymptotic stationarity is a very slight departure
from stationarity, Section~\ref{introrobustness} demonstrates that it
can have quite a significant impact on inference.
% 1.1.2 and motivated our new modeling ideas. %For our applications, we
                                %choose $\bG^{(T)}_0$ to correspond to
                                %an asymptotically stationary time series.
It is also important to observe  that although the process may be
stationary or asymptotically stationary
under $\bG^{(\bTau)}_0$, the same process when conditioned on
$\bG^{(\bTau)}$ is not even asymptotically stationary. Specifically,
\[
E \bigl(\gamma_t\mid\bG^{(\bTau)} \bigr)=\sum_{k=1}^{\infty}p_k\gamma^*_{kt},\qquad
\operatorname{Var}\bigl (\gamma_t\mid\bG^{(\bTau)} \bigr)=\sum_{k=1}^{\infty}p_k (\gamma^*_{kt} )^2
- \Biggl(\sum_{k=1}^{\infty}p_k\gamma^*_{kt} \Biggr)^2
\]
  and
\[
\operatorname{Cov} \bigl(\gamma_s,\gamma_t\mid\bG^{(\bTau)} \bigr)=\sum_{k=1}^{\infty}p_k\gamma^*_{ks}\gamma^*_{kt}
- \Biggl(\sum_{k=1}^{\infty}p_k\gamma^*_{ks} \Biggr)\Biggl (\sum_{k=1}^{\infty}p_k\gamma^*_{kt} \Biggr).
\]
Thus, $\bG^{(\bTau)}$ is nonstationary, although under $\bG^{(\bTau)}_0$,
$\bGamma_{\bTau}$ may have a stationary model so that the mean is constant and
the covariance depends  upon time only through the time lag
$|t-s|$. Thus, we have defined here a process $\bG^{(\bTau)}$ that is centered
around a stationary process, but is
actually nonstationary.
For purposes of applications, we have
given (ordered) time-points $(t_1,\ldots,t_T)$,   a
$T$-variate distribution $\bG^{(T)}$
on the space of all $T$-variate distributions of
$(\gamma_1,\ldots,\gamma_T)'$
with mean $\bG^{(T)}_0$ being the $T$-variate distribution implied
by a standard time series.

The
development of our nonstationary temporal process here technically
resembles that of a similar spatial process in \citet{Gelfand05}, but
%unlike in the case of the latter's spatial process,
differs from the latter in that it is actually embedded in the
model for the observed fMRI signals. As a result,
%matters and in fact, compared to the case of \ctn{Gelfand05}
the full conditional distributions of $\gamma_{ij}(t)$'s in our model
are much more general and %perhaps,
complicated than similar derivations
following \citet{Gelfand05}. Another important difference between our
approach and that of
\citet{Gelfand05} is that the latter had to introduce a pure error
(``nugget'') process to avoid discreteness of the distribution of
their spatial data. Such discreteness of the distribution (of our
temporal data) is naturally avoided here, however, owing to the embedding
approach used in our modeling. \citet{Gelfand05} also rely on the
availability of replications of the spatial dataset: our
embedding approach obviates this requirement by merely assuming
%requirement of replicated data with the %more realistic assumption of
the availability of replicated (unobserved) random
processes. We now introduce our dynamic effective
connectivity model.

%s2.2 ###
\subsection{A Dirichlet process-based dynamic effective connectivity
  model}
\label{secdynamicdp}
%s2.2.1 ###
\subsubsection{Hierarchical modeling}
For $i,j=1,2,\ldots, R$, define the $T$-compo\-nent vectors
$\bGamma_{ij}= (\gamma_{ij}(1),\gamma_{ij}(2),\ldots,\gamma_{ij}(T) )'$.
Further, let $\bGamma_{ij}$'s be %independent identically distributed
                                %with cumulative distribution $\bG$
                                %where
i.i.d. $\bG$, where $\bG
\sim \operatorname{DP}(\tau\bG_0)$, with  $\tau$ denoting the scale parameter quantifying
uncertainty in the base prior distribution $\bG_0$. Also, assume
that under\vspace*{-1pt} $\bG_0$, $\gamma_{ij}(1)\sim
N(\bar\gamma,\sigma^2_{\gamma})
$ and for
$t=2,\ldots,T$, $\gamma_{ij}(t)=\rho\gamma_{ij}(t-1)+\delta_{ij}(t)$,
where $|\rho|<1$ and $\delta_{ij}(t)\sim N(0,\sigma^2_{\delta})$ are
i.i.d. for $i,j=1,2,\ldots, R; t=1,2,\ldots,T$. It follows that under
$\bG_0$,
$\bGamma_{ij} \sim N_T(\bar\gamma\bmu_T, \bSigma)$ where
$\bmu_T=(1,\rho,\rho^2,\ldots,\rho^{T-1})'$ and for $s\leq t$,
$\bSigma$ has the $(s,t)$th element %under $\bG_0$ is given by
\begin{equation}
\Sigma_{st}=\rho^{s+t-2}\sigma^2_{\gamma}+\rho^{t-s}\sigma^2_{\delta} \biggl(\frac{1-\rho^{2(s-1)}}{1-\rho^2} \biggr).
\label{eqcovmatrix}
\end{equation}
Note that with $\bG_0$ as described above, the process is stationary
if we choose $\bar\gamma=0$ and
$\sigma^2_{\gamma}=\sigma^2_{\delta}/(1-\rho^2)$, otherwise
the process converges to stationarity for large~$s$. In other words,
under $\bG_0$, $E (\gamma_{ij}(s) ) =
E (\rho^{s-1}\gamma_{ij}(1)+\sum_{r=0}^{s-2}\rho^r\delta_{ij}(s-r) )
= \rho^{s-1}\bar\gamma$ which converges to 0 as $s\rightarrow\infty$,
while from (\ref{eqcovmatrix}) it follows that, as $s\rightarrow\infty$ with
$t-s<\infty$, $\Sigma_{st}\rightarrow\rho^{t-s}\sigma^2_{\delta}/(1-\rho^2)$.
The case for $s>t$ is similar.
Using the above developments, we specify our dynamic effective
connectivity model hierarchically, by augmenting (\ref{eqhier1}) and
(\ref{eqhier2}) with the following model for $\gamma_{ij}(t)$'s:
\[
\bGamma_{ij} \stackrel{\mathrm{i.i.d.}}{\sim}
\bG^{(T)}  \qquad \mbox{for } i,j=1,2,\ldots, R,  \mbox{ where }
\bG^{(T)}\sim \operatorname{DP}\bigl(\tau\bG^{(T)}_0\bigr).
\]
Distributional assumptions on $\epsilon_i(t)$'s, $\omega_i(t)$'s and
$\delta_{ij}(t)$'s are as in %\ctn{Bhattacharya06} on in
Section~\ref{subsecnimodel}. We use $\mM_{\mathrm{DP}}$ to refer to this
model: note also that  as
$\tau\rightarrow\infty$, our DP-based model converges to the AR(1)
model, which we denote using $\mM_{\mathrm{AR}}$. We note in closing that the
effective connectivity parameters are AR(1), hence VAR, under
the expected distribution of $\mM_{\mathrm{DP}}$. Of course, they are trivially
also so under $\mM_{\mathrm{AR}}$. Note, however, that given a realization of a
random distribution from the Dirichlet process, such VAR
representation does not hold.

%s2.2.2 ###
\subsubsection{Other prior distributions}
We specify independent prior distributions on each of
$\sigma^2_{\epsilon}$,$\sigma^2_w$,$\sigma^2_{\delta}$,$\rho$,$\tau$,
$\alpha_i$,$\beta_i(1)$ and
$\gamma_{ij}(1);i,j=1,2,\ldots, R$.
% (recall that $\tau$ is the scale parameter quantifying the strength
% of the belief we have on the prior base measure $\bG^{(T)}_0$)  as follows:
Specifically, $\alpha_i$'s are assumed to be i.i.d.
$N(\mu_i, \sigma^2_{\alpha})$ for $i=1,2,\ldots, R$ and
$\beta_i$'s are assumed to be i.i.d. $N(\bar\beta, \sigma^2_\beta)$,
for $i=1,2,\ldots, R$. Also, $\gamma_{ij}(1)$'s are independently
distributed with mean $\bar\gamma$ and variance $\sigma^2_{\gamma}$,
while  $\rho$ is  uniformly distributed in $(-1, 1)$, $\tau\sim
\Gamma(a_\tau,b_\tau)$ and $\sigma^{-2}_{\epsilon}$, $\sigma^{-2}_w$
and $\sigma^{-2}_{\delta}$ are each
i.i.d. %inverse Gamma
$\Gamma(a,b)$ with density\vspace*{1pt} having the functional form.
% $[\sigma^2_{\delta}]\propto
% \sigma_{\delta}^{-(b+2)}\exp(-a/{2\sigma^2_{\delta}})$.
%$\sigma^{-(b+2)}\exp(-a/{2\sigma^2})$.
Here $\mu_i,\sigma^2_{\alpha},\bar\beta$
and $\sigma^2_{\beta}$, $\bar\gamma$ and $\sigma^2_{\gamma}$, $a$,
$b$, $a_{\tau}$ and $b_{\tau}$ are all hyperparameters.
In our examples, we take $a=b=0$, reflecting our ignorance of
the unknown parameter $\sigma^2_{\delta}$. Although the Gamma priors
with $a=b=0$ are improper, they yielded proper posteriors in our case, vindicated
by fast convergence of the corresponding marginal chains and resulting right-skewed
posterior density estimates, which are expected of proper posteriors having positive support.
%({\bf Why? Is this because they are
%  noninformative? \color{red} Yes.})
%The parameters of the prior distribution of $\tau$ is chosen in the following manner:
For $(a_\tau,b_\tau)$ we first fix the expected value of
$\Gamma(a_{\tau},b_{\tau})$ (given by $a_{\tau}/b_{\tau}$) to be
such that in the full conditional distribution of $\bGamma_{ij}$, given by
(\ref{eqpriorcond}), the ``expected'' probability
of simulating a new realization from the ``prior'' base measure\vspace*{1pt}
approximately equals the probability
of selecting realizations of $\bGamma_{i'j'}$, for some $(i',j')\neq (i,j)$.
Hence, if there are $R^2$ nonzero $\bGamma_{ij}$ in the model, then setting
$a_{\tau}=c(R^2-1)$ and $b_{\tau}=c$ serves the purpose.
The resulting prior distribution has variance equal to its expectation if $c=1$.
To achieve large variance, we set $c=0.1$; the associated prior worked well
in our examples. We also experimented with $c=0.01$ and
$c=0.001$ and noted that while the case with $c=0.1$ provided the best
results (see Tables S-1 and S-2), inferences
related to the posterior distributions of the observed data were fairly
robust with respect to different choices of $c$. Moreover, the results
demonstrate that in terms of percentage of inclusion of the true
$\gamma_{ij}$'s,
% barring the inclusion percentages of  $\gamma_{32}$ and $\gamma_{33}$,
all inclusion percentages, with the exception of $\gamma_{32}$ and
$\gamma_{33}$, were quite robust with respect to $c$. The results
corresponding to $c=0.01$ and $c=0.001$ were  quite similar,\vspace*{-1pt} while
those corresponding to $c=0.1$ yielded better performance.
%, which we
%justify on  grounds of simplicity and also because we do not
%expect any over-dispersion   a priori.
Further, other hyperparameters were estimated empirically
from the data as in  \citet{Bhattacharya06} %employing
                                % \ctn{Bhattacharya06}'s
                                %computationally
              %efficient
using \citeauthor{Berger85}'s (\citeyear{Berger85}) ML-II approach.

%s2.2.3 ###
\subsubsection{Full conditional distributions}
The posterior distribution of the parameters are specified by
their full conditionals, which are needed for Gibbs sampling. The full
conditional distributions of $\alpha_i$, $\beta_i(t)$,
$\sigma^2_{\epsilon}$ and $\sigma^2_\omega$ are of
standard form (see Section S-1.1), while those of
the $\bGamma_{ij}$'s require some careful derivation. To describe these,
note that, on integrating out~$\bG^{(T)}$, the   prior  conditional
distribution of $\bGamma_{ij}$ given $\bGamma_{k\ell}$ for
$(k,\ell)\neq (i,j)$ follows a~Polya urn scheme, and is given by
\begin{equation}
%[\bGamma_{ij}\mid\bGamma_{k\ell};(k,\ell)\neq(i,j)]
%+\frac{\sum_{(k,\ell)\neq
%(i,j)}\delta_{\bGamma_{k\ell}}}{\tau+\#\{(k,\ell):(k,\ell)\neq
%(i,j)\}}.
[\bGamma_{ij}\mid\bGamma_{k\ell};(k,\ell)\neq(i,j)]\sim\frac{\tau\bG^{(T)}_0+{\sum_{(k,\ell)\neq
      (i,j)}\delta_{\bGamma_{k\ell}}}}{\tau+\#\{(k,\ell)\dvtx (k,\ell)\neq (i,j)\}}.
\label{eqpriorcond}
\end{equation}
The above Polya urn scheme shows that marginalization with respect to $\bG$ induces dependence
among $\bGamma_{ij}$ in the form of clusterings, while maintaining the same stationary marginal $\bG^{(T)}_0$
for each $\bGamma_{ij}$.
For Gibbs sampling we need to combine (\ref{eqpriorcond}) with the
rest of the model to obtain the \textit{full conditional}
distribution given all the other parameters and the data.
We obtain the full conditionals by first defining, for
$i,j=1,2,\ldots,R$, diagonal matrices
$\bA_{ij}=\sigma^{-2}_\omega\operatorname{diag}\{0,x^2(1)\beta^2_j(1),x^2(1)\beta^2_j(2),
\ldots,x^2(T-1)\beta^2_j(T-1)\}$, where\vspace*{1pt} $\operatorname{diag}$ lists the diagonal
elements of the relevant matrix.
We also define $T$-variate vectors~%
$\bB_{ij}$ for $i,j=1,2,\ldots,R$ with first element equal to
zero. For $t=2,\ldots,T$ the $t$th element of $\bB_{ij}$
is $B_{ij}(t) = \sigma_\omega^{-2}[\beta_i(t)\beta_j(t-1)x(t-1) -
\beta_j(t-1)x^{2}(t-1) \sum_{\ell=1\dvtx \ell \neq j}^R\gamma_{i\ell}(t) \beta_\ell(t-1)]$.
%=\frac{1}{\sigma^2_w} (\begin{array}{c}0\\\beta_1(1)\beta_2(1)\gamma_{12}(2)x^2(1)
%+\beta_1(1)\beta_3(1)\gamma_{13}(2)x^2(1)-\beta_1(2)\beta_1(1)x(1)\\
%-\beta_1(T)\beta_1(T-1)x(T-1)\\
Further, we note that, thanks to conditional independence, it is only
necessary to combine (\ref{eqpriorcond})
with (\ref{eqhier2}) to obtain the required full conditionals. It follows that
\begin{equation}
[\bGamma_{ij}\mid\cdots]\sim q^{(ij)}_0\bG^{(T)}_{ij}+\sum_{(k,\ell)\neq (i,j)}q^{(k\ell)}\delta_{\bGamma_{k\ell}},
\label{eqfullcond}
\end{equation}
where $\bG^{(T)}_{ij}$ is the $T$-variate normal distribution
with mean %$E (\bGamma_{ij}\mid\bG^{(T)}_{ij} )=
$(\bSigma^{-1}+\break\bA_{ij})^{-1}
(\bar\gamma\bSigma^{-1}\times \bmu+\bB_{ij})$ %(with $\bone_T$ being a
                                %$T$-variate vector of ones)
and variance
%$Var (\bGamma_{ij}\mid\bG^{(T)}_{ij} )=
$(\bSigma^{-1}+\bA_{ij})^{-1}$.
Also,
\begin{eqnarray}
\label{eqq0}
%q^{(ij)}_0&\propto&\frac{\tau}{|\bSigma|^{\frac{1}{2}}|\bSigma^{-1}+\bA_{ij}|^{\frac{1}{2}}}\nonumber\\
%&\times&\exp [-\frac{1}{2} \{{\bar\gamma}^2\bmu_T'\bSigma^{-1}\bmu_T
%  -(\bar\gamma\bSigma^{-1}\bmu_T+\bB_{ij})'(\bSigma^{-1}+\bA_{ij})^{-1}(\bar\gamma\bSigma^{-1}\bmu_T+\bB_{ij}) \} ]\nonumber\\
q^{(ij)}_0 %\propto
 &= & C\frac{\tau}{|\bI+\bSigma\bA_{ij}|^{\fraca{1}{2}}}\nonumber\\
 &&{}\times
\exp \biggl[-\frac{1}{2}\{\bar\gamma^2\bmu_T'\bSigma^{-1}\bmu_T
\\
&&\hphantom{{}\times
\exp \biggl[-\frac{1}{2}\{}{}
  -(\bar\gamma\bSigma^{-1}\bmu_T+\bB_{ij})'(\bSigma^{-1}+\bA_{ij})^{-1}(\bar\gamma\bSigma^{-1}\bmu_T+\bB_{ij})\}
  \biggr]
\nonumber
\end{eqnarray}
  and
\begin{equation}
q^{(k\ell)}
=C\exp \bigl[-\tfrac{1}{2}(\bGamma_{k\ell}-\bA^{-1}_{ij}\bB_{ij})'\bA_{ij}(\bGamma_{k\ell}
-\bA^{-1}_{ij}\bB_{ij})-\bB_{ij}'\bA^{-1}_{ij}\bB_{ij} \bigr]
\label{eqq1}
\end{equation}
 for $ (k,\ell)\neq (i,j)$, with $C$ chosen to satisfy
$ q^{(ij)}_0+\sum_{(k,\ell)\neq (i,j)}q^{(k\ell)}=1$.\vspace*{1pt}
Observe that unlike all DP-based
approaches hitherto considered in the statistics literature, in our case
$\bG^{(T)}_{ij}$, the conditional posterior base measure is not
independent of $\bGamma_{i'j'}$ for $(i',j')\neq (i,j)$, which is a consequence
of the fact that, thanks to (\ref{eqhier2}), $\bGamma_{i'j'}$ are not conditionally
independent of each other. Thus, our methodology generalizes other DP-based methods,
including that of \citet{Gelfand05}.

Section S-1.2 presents an alternative algorithm to updating
$\bGamma_{ij}$ using \textit{configuration indicators} which are
updated sequentially using themselves and only the distinct
$\bGamma_{ij}$, given everything else.
%configuration indicators theoretically improves convergence
%properties of the Markov chain:
\citet{MacEachern94} has argued that such an updating procedure
theoretically improves convergence properties of the Markov chain:
however, Section S-1.3 shows that in our case the associated
conditional  distributions need to be obtained separately for each of
the $2^9$ possible configuration indicators. This being infeasible, we
recommend (\ref{eqfullcond}) for updating $\bGamma_{ij}$.
[We remark here that full conditionals are easily obtained using
configuration indicators in the case of
\citet{Gelfand05}, thanks to the
relative simplicity of their spatial problem.] Also, (\ref{eqq0}) and
(\ref{eqq1}) imply that as $\tau\rightarrow\infty$,
the full conditional distribution (\ref{eqfullcond})
converges to $\bG^{(T)}_{ij}$, which\vspace*{1pt} is actually the full conditional
distribution of the entire $T$-dimensional parameter vector
$\bGamma_{ij}$ under the AR(1) model. In either case, we
provide computationally efficient multivariate updates for our Gibbs
updates: this makes our problem computationally tractable.

To obtain  the full conditional of $\tau$, define
$m=\#\{(i,j);i,j=1,2,\ldots, R\}=R^2$. Then, note that as in
\citet{Escobar95}, for a $\Gamma(a_{\tau},b_{\tau})$ prior
on~$\tau$, the full conditional distribution of the latter, given the
number ($d$) of distinct $\bGamma_{ij}$ %, denoted by $d$,
and another continuous random variable $\eta$, is a mixture of two
$\operatorname{Gamma}$ distributions, specifically
$\pi_{\eta}\Gamma(a_{\tau}+d,b_{\tau}-\log(\eta))\nonumber +
(1-\pi_{\eta})\Gamma(a_{\tau}+d-1,b_{\tau}-\log(\eta))$,
where
${\pi_{\eta}}/({1-\pi_{\eta}})=({a_{\tau}+d-1})/({m(b_{\tau}-\log(\eta))})$.
Also, the full conditional of $\eta$ is $\upbeta(\tau+1,m).$ %  i.e.,
                                % a $Beta$ distribution with mean
                                % $(\tau+1)/(\tau+m+1)$.
Finally, the full conditional distributions of $\sigma^2_{\delta}$ and
$\rho$ are
not %available in closed form
very standard and need careful derivation. Section~S-1.4 describes a
Gibbs sampling approach using configuration sets for updating
$\sigma_\delta$ and $\rho$. For implementing\vadjust{\goodbreak} this Gibbs step, one does
not need to simulate the configuration indicators, as they can be
determined after simulating
the $\bGamma_{ij}$'s using (\ref{eqfullcond}). Hence, this step is feasible.
However, we
failed to achieve sufficiently good convergence with this approach, and
hence used a~Metropolis--Hastings step.
The acceptance ratio for the Metropolis--Hastings step is
given by
$[\bGamma_{11}][\bGamma_{12}\mid\bGamma_{11}][\bGamma_{13}\mid\bGamma_{12},\bGamma_{11}]\cdots
[\bGamma_{33}\mid\bGamma_{32},\ldots,\bGamma_{11}]$, evaluated,
respectively, at the new and the old values of the\vspace*{-1pt} parameters
$(\sigma^2_{\delta},\rho)$. In the above, $[\bGamma_{11}]\sim\bG^{(T)}_0$,
and the other factors are Polya urn distributions, following
easily from (\ref{eqpriorcond}). Once again, note the use of
multivariate updates in the MCMC steps, making our updating approach
computationally feasible and easily implemented.

We conclude this section by noting that our model is structured to be
identifiable. The priors of $\alpha_i$,  $\beta_i(t)$,
$\gamma_{ij}(t)$ are all different and informative. Further,
(\ref{eqhier2}) shows that $\beta_i(t)$ is not permutation-invariant
with respect to the indices of $\bGamma_{ij}$'s. %The variance
                                %parameters all have the same
                                %(improper) prior
Identifiability of our model is further supported by the results in
this paper, which show all posteriors (based on MCMC) to be distinct
and different. This is unlike the case of the usual Dirichlet
process-based  mixture models which are permutation-invariant, as in
\citet{Escobar95}, where the parameters have the same posterior due to
nonidentifiability. We now investigate performance of our methodology.
%s3 ###
\section{Simulation studies}
\label{secsimstudy}
%Before describing our simulation method, it is important to note that
%$\gamma_{31}(t)=\gamma_{32}(t)=0$ for all $t$, actually outperforms other models,
%and that the assumption is tenable.
%The interpretation is that there was no feedback from the two sites of
%control (LG and MOG) to the source of control (DLPFC) directly.

%In our set up, from a technical point of view,
%the restriction $\gamma_{31}(t)=\gamma_{32}(t)=0$ has the effect of breaking
%nonidentifiability problems in our model.
We performed a range of simulation experiments to investigate
performance of our approach relative to its alternatives. Since there
are 9 nonzero $\bGamma_{ij}$'s in our model, we followed the recipe
provided in Section \ref{secmethodology} and  put a $\Gamma(0.8,0.1)$
prior on the DP scale parameter $\tau$. We investigated fitting
$\mM_{\mathrm{DP}}$, $\mM_{\mathrm{AR}}$ and $\mM_{\mathrm{RW}}$ to the simulated data of Section
\ref{introrobustness} and also to data simulated from the $\mM_{\mathrm{RW}}$
and $\mM_{\mathrm{AR}}$ models, the latter with both $\rho = 0.5$ (clearly
stationary model) and $\rho=0.95$ (where the model is not so
clearly distinguished from nonstationarity but more clearly
distinguished than when $\rho=0.999$). The Gibbs sampling procedure
for model $\mM_{\mathrm{AR}}$ in our simulations was very similar to that
of the $\mM_{\mathrm{RW}}$ detailed in \citet{Bhattacharya06}: we
omit details. For all experiments in this paper and in the supplement, we
discarded the first 10,000 MCMC iterations as burn-in and stored
the following 20,000 iterations for Bayesian inference. Our
results are summarized here for want of space, but presented in detail
in Section S-2, with performance evaluated graphically [in terms of
the posterior densities of $\gamma_{ij}(t)$'s] and numerically using
coverage and average lengths of the 95\% HPD credible intervals of the
posterior predictive distributions (for details, see Section S-2).

The results of our experiments using the simulated data of
Section~\ref{introrobustness} showed that $\mM_{\mathrm{AR}}$
performed better than $\mM_{\mathrm{RW}}$, but model $\mM_{\mathrm{DP}}$ was the clear
winner. Indeed, the support of the posterior distributions of
$\gamma_{22}(t)$ and $\gamma_{23}(t)$ using $\mM_{\mathrm{AR}}$ were much too wide
to be of much use, but substantially narrower under
$\mM_{\mathrm{DP}}$. $\mM_{\mathrm{DP}}$ also outperformed the other two models in terms
of the proportion of true $\gamma_{ij}(t)$'s included in the
corresponding 95\% HPD CIs. %The posterior of $\rho$ under $\mM_{DP}$
                            %also had much wider support than that
                            %using $\mM_{AR}$. Additionally,
These CIs also captured almost all of the true values of
$\gamma_{ij}(t)$ under $\mM_{\mathrm{DP}}$, but far fewer values using
$\mM_{\mathrm{AR}}$.  $\mM_{\mathrm{DP}}$ also exhibited better predictive performance
than $\mM_{\mathrm{AR}}$ and $\mM_{\mathrm{RW}}$. All these findings which favor our DP-based model were implicitly the consequence of the fact that
the true model in our experiment was approximately nonstationary, and
modeled more flexibly by our nonstationary DP model rather than the
stationary AR(1) model.
That this borderline between stationarity and nonstationarity of the
true model is important was vindicated by the results of fitting
$\mM_{\mathrm{RW}}$, $\mM_{\mathrm{AR}}$ and $\mM_{\mathrm{DP}}$ on the dataset simulated using
$\mM_{\mathrm{RW}}$. Here, $\mM_{\mathrm{RW}}$ outperformed both $\mM_{\mathrm{DP}}$
and $\mM_{\mathrm{AR}}$ in terms of coverage of the true values of~%
$\gamma_{ij}(t)$, indicating that $\mM_{\mathrm{DP}}$ may under-perform when
compared to the true
model, in terms of coverage of parameter values, when the  true model
can be clearly identified. In terms of prediction ability, however,
$\mM_{\mathrm{DP}}$ was still the best performer, with the best coverage of the
data points by the posterior predictive distribution and the
lengths of the associated 95\% CIs. This finding was not unexpected,
since $\mM_{\mathrm{DP}}$ involves  model averaging (see Section S-1.5), which
improves predictive performance  [see, e.g., \citet{Kass95}].
For the dataset simulated from $\mM_{\mathrm{AR}}$ with  $\rho=0.5$, the
true model ($\mM_{\mathrm{AR}}$) outperformed $\mM_{\mathrm{DP}}$ marginally and
$\mM_{\mathrm{RW}}$ substantially, but when $\rho=0.95$,
$\mM_{\mathrm{DP}}$ provided a much better fit than  $\mM_{\mathrm{AR}}$ or
$\mM_{\mathrm{RW}}$. We have already mentioned that
$\mM_{\mathrm{DP}}$ outperformed $\mM_{\mathrm{AR}}$ (and $\mM_{\mathrm{RW}}$) for the borderline
case of  $\rho=0.999$:
the experiment with $\rho=0.95$ demonstrated good performance of
$\mM_{\mathrm{DP}}$ even in relatively more distinguishable situations. At the
same time, the experiment with $\rho=0.5$ warns against over-optimism
regarding $\mM_{\mathrm{DP}}$; for clearly stationary data, we are at least
marginally better off replacing $\mM_{\mathrm{DP}}$ with a stationary model
such as $\mM_{\mathrm{AR}}$. In spite of this caveat for clearly stationary
situations, our simulation experiments indicated that our
DP-based approach is flexible enough to address stationary models
as well as deviations. We now analyze the Stroop
Task dataset introduced in Section~\ref{introACN}.
%%%%%%%%%%%%%%%%%%%%%%%%%%%%%%%%%%%%%%%%%%%%%%%%%%%%%%%%%%%%%%%%%%%%%%%%%%%%%%%%%%%%%%%%%%%%%%%%%%%%%
%
%indicates the true value of $\rho$.}
%The vertical line indicates the true value of $\rho$.}

%s4 ###
\section{Application to Stroop task data}
\label{secreal}
The dataset was preprocessed following \citet{Ho05} and
Bhattacharya, Ho and  Pur\-kayastha \citeyear{Bhattacharya06}, to which we refer for details while
providing only a brief summary here. For each of the three regions
(LG, MOG and DLPFC), a~spherical region of 33 voxels was
drawn around the location of peak activation.
The voxel-wise time series of the selected voxels in
each region were then subjected to higher order (multi-linear)
singular value decomposition (HOSVD) using methods in
 \citet{lathauweretal00}. The first mode of this HOSVD, after detrending
with a running-line smoother as in \citet{marchiniandripley00}, provided
us with our detrended time series response $y_i(t)$ for the $i$th
region [see Figure~S-4 for $y(t)$'s as well as $x(t)$].

%f2 ###
\begin{figure}

\includegraphics{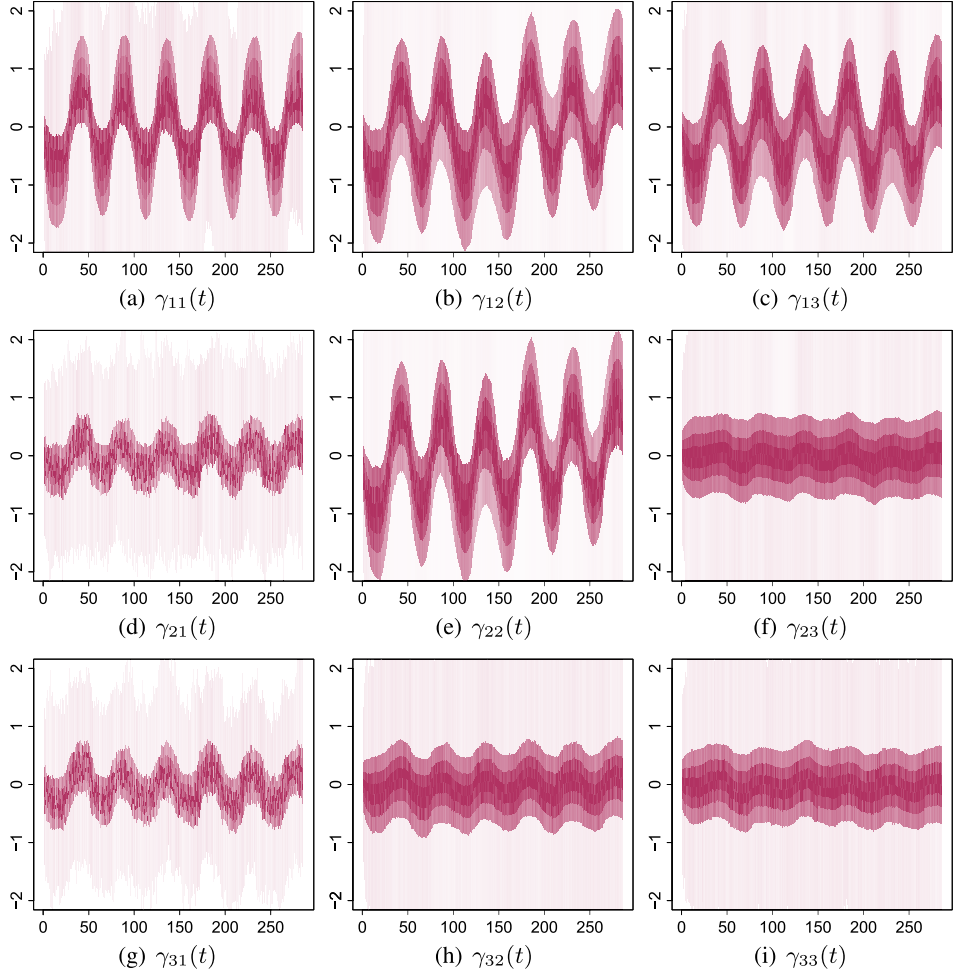}

\caption{Estimated posterior densities (means in solid lines) of the regional influences over
  time.}
\label{figdpunrestricted}
\end{figure}

We compared results obtained using $\mM_{\mathrm{DP}}$ with those using
$\mathcal M_{\mathrm{RW}}$\break and $\mathcal M_{\mathrm{AR}}$. We refer
to \citet{Bhattacharya06} and the supplement for detailed results using
$\mathcal M_{\mathrm{RW}}$  and $\mathcal M_{\mathrm{AR}}$, respectively, only
summarizing them here in comparison with results obtained using
$\mathcal M_{\mathrm{DP}}$, which we also discuss in greater detail here.
%We also repeated our calculations with model $\bM_3$, unrestricted
%case, that is $\bM_3$ without the restriction
%$\gamma_{31}(t)=\gamma_{32}(t)=0$. Figure
%(\ref{figrealdatadpunrestricted}) shows the densities of the
%corresponding posterior distributions of $\gamma_{ij}(t)$,
%represented by 8 quantiles. The posterior distributions do not
%exhibit any clear oscillatory structures, except in the case of
%$\bGamma_{11}$. Experiments with model $\bM_2$ (see Figures
%versions respectively) confirm that model $\bM_3$ with the condition
%$\gamma_{31}(t)=\gamma_{32}(t)=0$ for all $t$ remains unbeaten.
%In all cases, we used Gibbs' sampling to obtain samples from the
%posterior distribution and discarded the first 10,000 iterations as
%burn-in, retaining the next 20,000 for inference.
Detailed studies on MCMC convergence are in Section S-3.2.

%f3 ###
\begin{figure}[b]
\vspace*{-1pt}
\includegraphics{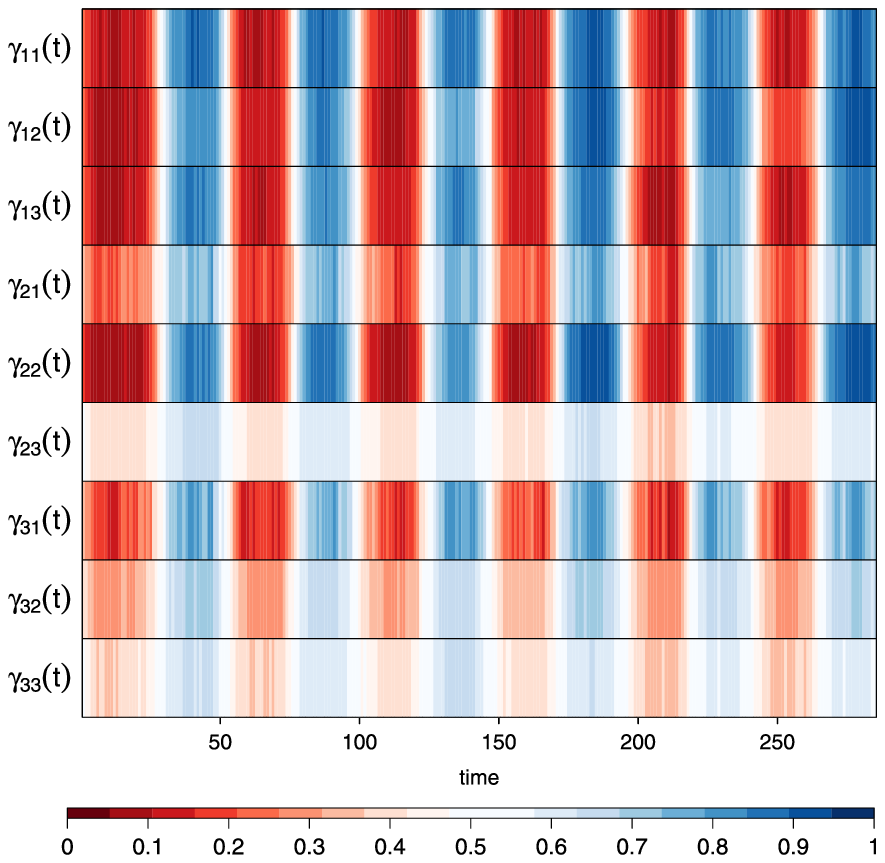}

\caption{Proportions of estimated marginal posterior density of
$\gamma_{ij}(t)$ with positive support at $t$.}
\label{figmap}
\end{figure}

%s4.1 ###
\subsection{Results}
\label{subsecresults}

Figure~\ref{figdpunrestricted} displays the Gibbs-estimated marginal
posterior distributions of the $\gamma_{ij}(t)$'s for each time point
$t$ obtained using $\mM_{\mathrm{DP}}$.
A~striking feature of the marginal posterior densities of
Figure~\ref{figdpunrestricted} is the very strong oscillatory nature
of these effective connectivity parameters with the modeled BOLD response
$x(t)$. This is quite different from the posterior distributions of
$\gamma_{ij}(t)$'s obtained using $\mM_{\mathrm{AR}}$ (see
Figure~S-7). Table~\ref{tablerealproportions} evaluates performance
of the two models in terms
%or each $t$ (on the horizontal axis), the figure displays the
%estimated marginal posterior density of $\gamma_{ij}(t)$ at that
%ime-point, with opacity representing denser regions of the density.
%osterior means over time are also displayed by means of the solid
%line.
of the length and proportion of
observations contained in the 95\% HPD credible intervals of the
posterior predictive distributions:
the intervals obtained using
$\mM_{\mathrm{DP}}$ have greater coverage but are also much narrower, making it
by far the better choice among the models.
Figure~\ref{figdpunrestricted} also shows that
$\gamma_{23}(t),\gamma_{32}(t)$ and $\gamma_{33}(t)$---and, to a
lesser extent, $\gamma_{21}(t)$ and $\gamma_{31}(t)$---oscillate differently from the others in that their
amplitude is close to zero. We examined this issue further through
%t2 ###
  \begin{table}
  \tabcolsep=0pt
  \tablewidth=265pt
\caption{Proportions of observed $y$  included in, and average length, of
 the 95\% credible
  intervals of the  posterior predictive distributions
  under $\mM_{\mathrm{AR}}$~and $\mM_{\mathrm{DP}}$  for the Stroop task dataset}
\label{tablerealproportions}
  \begin{tabular*}{265pt}{@{\extracolsep{\fill}}lcccc@{}}\hline
   & \multicolumn{2}{c}{\textbf{Proportions}}& \multicolumn{2}{c@{}}{\textbf{Average length}} \\[-5pt]
 & \multicolumn{2}{c}{\hrulefill}& \multicolumn{2}{c@{}}{\hrulefill}\\
     $\mathbf{y}$  & $\boldmathcaligr M_{\mathrm{\mathbf{AR}}}$ & $\boldmathcaligr M_{\mathrm{\mathbf{DP}}}$ & $\boldmathcaligr M_{\mathrm{\mathbf{AR}}}$ & $\boldmathcaligr M_{\mathrm{\mathbf{DP}}}$\\ \hline
    $y_1$  & 0.92     & 0.99   & 4,960.9     & 2,215.1\\
    $y_2$  & 1.00     & 1.00   & 3,864.2     & 2,068.1\\
    $y_3$  & 1.00     & 1.00   & 4,352.8     & 2,084.3\\
    \hline
  \end{tabular*}
  \vspace*{-3pt}
\end{table}
Figure~\ref{figmap}, which provides a map of the proportions of the
cases for which each estimated marginal posterior density of
$\gamma_{ij}(t)$ has positive  support at ti\-me~$t$. The intensities are
mapped via a red-blue
diverging palette: thus, darker hues of
 blue and red indicate high and low
values, respectively, for the proportions. Lighter hues of red or blue
indicate values in the middle.
Clearly, very little proportion of the marginal density  is either on
the positive or the negative parts of the
real line for the cases of $\gamma_{23}(t)$, $\gamma_{32}(t)$ and~$\gamma_{33}(t)$.
We therefore investigated performance of models $\mM_{\mathrm{DP}}$ modified
to exclude some or all of these regional influences.

%s4.1.1 ###
\subsubsection{Investigating restricted submodels of $\mM_{\mathrm{DP}}$}
\label{subsubsecsubmodels}
Bhattacharya, Ho and  Purkayastha (\citeyear{Bhattacharya06}) found that the model ${\mathcal M}_{\mathrm{RW}}$ with the
constraint $\gamma_{31}(t)=\gamma_{32}(t) = 0$ (henceforth ${\mathcal
  M}^-_{\mathrm{RW}}$) provided better results that the unconstrained
%(left columns) and average length of (right columns)   the 95\%
%credible intervals of the posterior predictive   distributions of
%$y_1,y_2,y_3$ for our    %restricted ($\mathcal M_{AR}^-$) and
%unrestricted AR(1) ($\mathcal M_{AR}$)    %and restricted ($\mathcal
%M_{AR}^-$)    and unrestricted DP-based ($\mathcal M_{DP}$) models.}
%& \multicolumn{2}{|c|}{Average length}
%$y$ & $\mathcal M_{AR}$ & $\mathcal M_{DP}$ & $\mathcal M_{AR}$ & $\mathcal M_{DP}$\\ \hline
%$y_1$  & 0.92     & 0.99   & 4,960.9     & 2,215.1\\
%$y_2$  & 1.00     & 1.00   & 3,864.2     & 2,068.1\\
%$y_3$  & 1.00     & 1.00   & 4,352.8     & 2,084.3\\
%%\vspace{-0.2in}
$\mM_{\mathrm{RW}}$. Figure~\ref{figdpunrestricted} also points to the
possibility that models with some $\gamma_{ij}(t)\equiv 0$ might
provide better performance. We explored these aspects
quantitatively using the models ${\mathcal
  M}_{\mathrm{AR}}$ and ${\mathcal M}_{\mathrm{DP}}$, by considering the proportion of
data contained in, and the average lengths of, the 95\% HPD CIs of the
corresponding posterior predictive distributions of
$y_i(t); i=1,2,3, t=1,\ldots,T$.
A~systematic evaluation of all possible submodels is computationally
very time-consuming, so we investigated models with combinations of
$\gamma_{31}(t)=\gamma_{32}(t)\equiv 0$ as in
\citet{Bhattacharya06} and with null $\gamma_{ij}(t)$'s for those $(i,j)$'s
whose posterior distributions exhibited   less amplitude of
oscillation as per Figure~\ref{figdpunrestricted}. Table~\ref{tablerealproportions2}
summarizes performances of the top three submodels: others
are in Tables S-11 and~S-12. The top three performers were the following:
\begin{itemize}
  \item $\mM_{\mathrm{DP}}^{(1)}$: $\mM_{\mathrm{DP}}$ but with $\gamma_{33}(t)\equiv 0
    \ \forall  t$.\vspace*{2pt}
  \item $\mM_{\mathrm{DP}}^{(2)}$: $\mM_{\mathrm{DP}}$ but with $\gamma_{32}(t)\equiv 0
  \ \forall  t$.\vspace*{3pt}
  \item $\mM_{\mathrm{DP}}^{(3)}$: $\mM_{\mathrm{DP}}$ but with
    $\gamma_{32}(t)=\gamma_{33}(t)\equiv 0 \ \forall  t$.
\end{itemize}
Thus, $\mM^{(1)}_{\mathrm{DP}}$ and $\mM_{\mathrm{DP}}^{(2)}$ both beat $\mM_{\mathrm{DP}}$
(of Table~\ref{tablerealproportions}). The average 95\% posterior
predictive length using $\mM_{\mathrm{DP}}^{(2)}$ is
about midway between $\mM_{\mathrm{DP}}^{(1)}$ and the unrestricted DP-based
model, so we report our final findings and conclusions
only using $\mM_{\mathrm{DP}}^{(1)}$.
\begin{table}[b]
\tabcolsep=0pt%
\caption{Proportions of the observed data in, and mean
  lengths of, the 95\% credible~intervals of   posterior predictive
  distributions of $y_1,y_2,y_3$ and the mean lengths of the 95\%
  credible intervals~for~the~top~three candidate submodels}
\label{tablerealproportions2}
  \begin{tabular*}{\textwidth}{@{\extracolsep{\fill}}lcccccc@{}}
  \hline
   & \multicolumn{3}{c}{\textbf{Proportion}} & \multicolumn{3}{c@{}}{\textbf{Mean length}}\\[-5pt]
 & \multicolumn{3}{c}{\hrulefill} & \multicolumn{3}{c@{}}{\hrulefill}\\
%    & $\mM^{(1)}_{DP}$ & $\mM^{(2)}_{DP}$ & $\mM^{(3)}_{DP}$ & $\mM^{(4)}_{DP}$ & $\mM^{(5)}_{DP}$
%    &  $\mM^{(1)}_{DP}$ & $\mM^{(2)}_{DP}$ & $\mM^{(3)}_{DP}$ & $\mM^{(4)}_{DP}$ &  $\mM_{DP}^{(5)}$ \\ \hline
%    $y_1$ & 0.93 & 0.99 & 1.0 & 1.0 & 0.98 & 4,946.1 & 3,824.2 & 3,238.8 & 3,241.4 & 3,856.8 \\
%    $y_2$ & 1.0 & 1.0 & 1.0 & 1.0 & 1.0 & 4,201.6 & 3,234.9 & 2,864.4
%    &  2,872.8 & 3,406.0 \\
%    $y_3$  & 1.0 & 1.0 & 1.0 & 1.0 & 1.0 & 4,438.8 & 3,594.8 &
%    2,980.1 &   2,953.3 & 3,485.9 \\
    $\mathbf{y}$ & $\boldmathcaligr M_{\mathbf{DP}}^{\bolds{(1)}}$ & $\boldmathcaligr M_{\mathbf{DP}}^{\bolds{(2)}}$
    & $\boldmathcaligr M_{\mathbf{DP}}^{\bolds{(3)}}$
    &  $\boldmathcaligr M_{\mathbf{DP}}^{\bolds{(1)}}$ & $\boldmathcaligr M_{\mathbf{DP}}^{\bolds{(2)}}$
    & $\boldmathcaligr M_{\mathbf{DP}}^{\bolds{(3)}}$
    \\ \hline
    $y_1$ & 0.99 & 0.99 & 1.0 & 2,097.6 & 2,140.4 & 2,276.5 \\
    $y_2$ & 1.0\hphantom{9} & 1.0\hphantom{9} & 1.0 & 1,971.6 & 2,019.5 & 2,127.8 \\
    $y_3$ & 1.0\hphantom{9} & 1.0\hphantom{9} & 1.0 & 1,985.0 & 2,021.3 & 2,125.4 \\ \hline
  \end{tabular*}
\end{table}

%f4 ###
\begin{figure}

\includegraphics{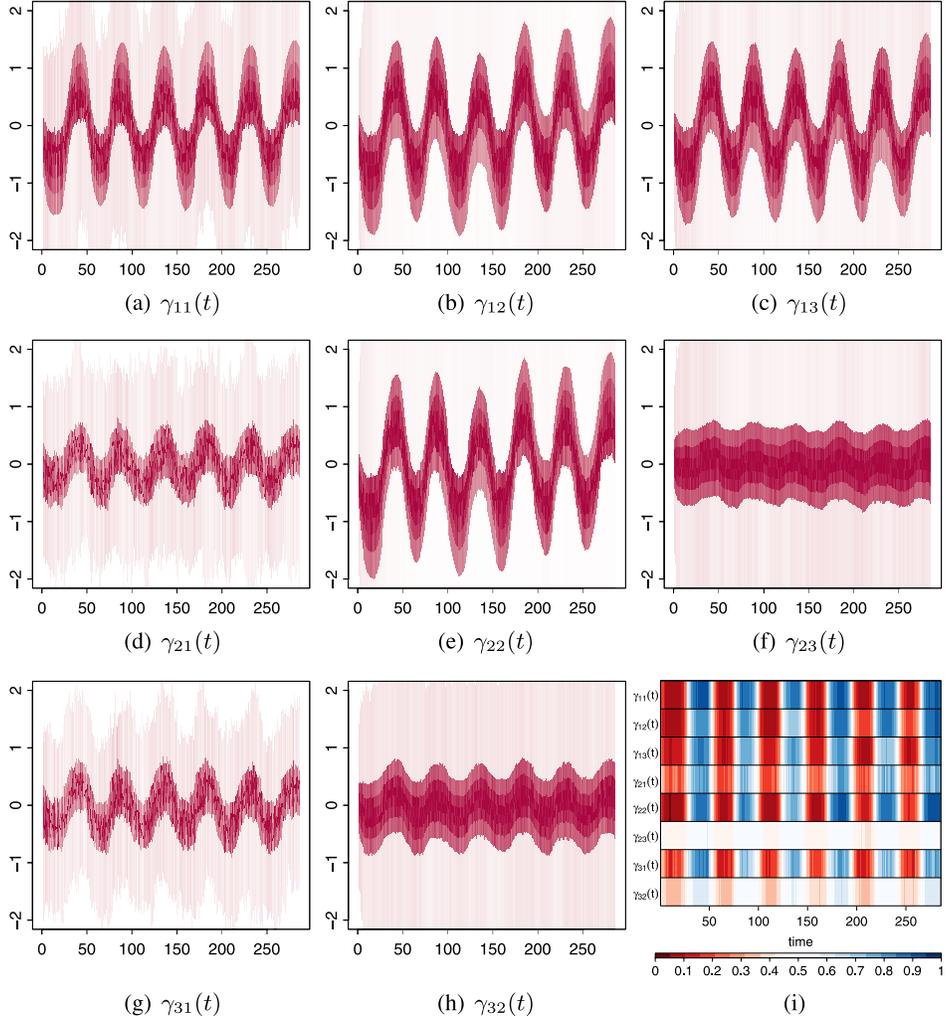}

\caption{\textup{(a)--(h)} Estimated posterior densities (means in solid lines) of the
nonnull regio\-nal influences over time using $\mM_{\mathrm{DP}}^{(1)}$.
\textup{(i)} Proportion of the posterior distribution of~$\gamma_{ij}(t)$ with positive support at time $t$.}
\label{figdpg330}
\end{figure}

%s4.2 ###
\subsection{Summary of findings}

Figure~\ref{figdpg330}(a)--(h) display the posterior densities of
the nonnull regional influences $\gamma_{ij}(t)$'s over time. These
$\gamma_{ij}(t)$'s are very similar to those in
Figure~\ref{figdpunrestricted}(a)--(h), with nonzero effective
connectivity parameters again having a very pronounced oscillation synchronous
with the modeled BOLD response: indeed, only the
$\gamma_{23}(t)$ of Figure~\ref{figdpg330}(f) has an oscillation
slightly more damped than in Figure~\ref{figdpunrestricted}.
% In fact, the correlation of $x(\cdot)$ with the posterior means  of
% $\gamma_{11}(\cdot)$, $\gamma_{12}(\cdot)$, $\gamma_{13}(\cdot)$,
% $\gamma_{21}(\cdot)$, $\gamma_{22}(\cdot)$, $\gamma_{23}(\cdot)$,
% $\gamma_{31}(\cdot)$, $\gamma_{32}(\cdot)$ and $\gamma_{33}(\cdot)$
% are 0.94, 0.84, 0.91, 0.92, 0.86, 0.79, 0.91, 0.88, and 0.70,
% respectively.  In fact, the correlations are much higher if a
% smoothed version of $x(\cdot)$ is considered. This is described in
% Section 3 of the supplementary document.
%
% The results show that the oscillations are strongly synchronous with
% the oscillations in the HRF $x(\cdot)$, and even more strongly
% synchronous with the smoothed HRF $\hat x(\cdot)$.
%
% Perhaps the oscillation with $x(t)$ of highest amplitude is seen in the  posterior density of $\gamma_{11}(t)$; %, $\gamma_{12}(t)$,
                                %$\gamma_{13}(t)$ and
                                %$\gamma_{22}(t)$:
Further, Figure~\ref{figdpg330}(i) indicates that the estimated
posterior densities put most of their mass either below zero [when $x(t)$ is
negative] or above zero [when $x(t)$ is positive]. Indeed, these
densities have  substantial mass around zero only when $x(t)$ is
around zero. We also smoothed the modeled BOLD response $x(t)$ to explore
further its relationship with each of the estimated posterior mean
$\gamma_{ij}(t)$'s from~$\mM_{\mathrm{DP}}^{(1)}$. For each $t$, we specified
$x(t) =A\cos(2\pi\omega t+\phi)+\psi_t$ where $\psi_t$ %\stackrel{iid}
are i.i.d.~$N(0,\sigma^2_{\psi})$, $A$ is the amplitude of the time
series, $\omega$ is the  oscillation frequency and $\phi$ is a phase
shift. Equivalently, $x(t)=\beta_1\cos(2\pi\omega
t)+\beta_2\sin(2\pi\omega t)+\psi_t$ with $\beta_1=A\cos(\phi)$ and
$\beta_2=A\sin(\phi)$. We obtain %$\omega$ to be
$\hat\omega = 0.02$
using the periodogram approach [see, i.e., \citet{Shumway06}].
Thus, each cycle in $x(t)$ has a length of about 50 time-points. A
least squares fit yields $\hat\beta_1 = 0.27$ and $\hat\beta_2 =
-0.61,$ hence, $\hat A=0.80$ and $\phi=1.16$. Figure~S-8 shows that the
smoothed BOLD response $\hat x(t)=\hat\beta_1\cos(2\pi\hat\omega
t)+\hat\beta_2\sin(2\pi\hat\omega t)$ closely approximates the
original time series $x(t)$. The correlation of $\hat x(t)$ with each
of $\gamma_{11}(t)$, $\gamma_{12}(t)$, $\gamma_{13}(t)$,
$\gamma_{21}(t)$, $\gamma_{22}(t)$, $\gamma_{23}(t)$, $\gamma_{31}(t)$
and $\gamma_{32}(t)$ are 0.959, 0.909, 0.952, 0.950, 0.922, 0.874,
0.949 and 0.929, respectively. Thus, $\gamma_{ij}(t)$'s are
not completely linear in the BOLD response, but very close to being so
with regard to its transformed version.

The results of our analysis indicate that the region LG, centered
around zero, exhibits very strong evidence of self-feedback,
oscillatory with high amplitude, and period of about 50, matching the
period of the modeled BOLD response $x(t)$. Similar influences are
exerted by both MOG and DLPFC on LG and by the MOG region on
itself. Indeed, Figure \ref{figdpg330} indicates that these four
inter- and intra-regional influences have, broadly, a similar pattern
in terms of amplitude. The influence of LG on MOG and DLPFC is smaller
and similar to each other. Further, Figure  \ref{figdpg330}(f) and (h)
indicate that the feedback provided by DLPFC on MOG [$\gamma_{23}(t)$]
is similar to that in the reverse direction [$\gamma_{32}(t)$]. Thus,
there are three broad patterns in the way that inter-and
intra-regional influences occur.
%A similar set of patterns is also provided in
%Figure~\ref{figdpunrestricted}, corresponding to the unrestricted
%model with the only difference being that  the third cluster  also
%includes $\bGamma_{33}$.
% Broadly, the result corresponding to $\mM^{(1)}_{DP}$ indicates that
% all the 3 regions inflence the region LG in the same way, in turn,
% LG also influences the other two regions MOG and DLPFC in a like
% manner. The same inflence is exerted by MOG and DLPFC on one
% another. The self-feedback of the region MOG is the same as the
% influence exerted by LG, MOG and DLPFC on LG. These are new
% findings, not reported elsewhere in the literature.

Our analysis also demonstrates the existence of the ACN and its
mechanism while performing a Stroop task. Thus, the executive control
system (DLPFC) provides instruction to both the task-irrelevant (LG)
and task-relevant processing sites (MOG) but gets similar levels of
feedback from the task-relevant processor (MOG). LG which sifts out
the task-irrelevant color information gets a lot of feedback in doing
so from both itself and MOG. However, it provides far less feedback
to the task-relevant shape information processing MOG  and the
executive control DLPFC. MOG itself provides substantial self-feedback
while processing shape information. Finally, note that while our
results indicate higher amplitudes for inter-regional feedback
involving $\gamma_{ij}(t)$'s when they involve LG rather than MOG, this
is consistent with the established notion that processing shape
information is a higher-level (more difficult) cognitive function than
distinguishing color.

The results on the effective connectivity parameters using ${\mathcal
  M}_{\mathrm{DP}}$ are very different from those done using ${\mathcal
  M}^{-}_{\mathrm{RW}}$ [see Figure~5 of  \citet{Bhattacharya06}] or
$\mathcal M_{\mathrm{AR}}$.  Using ${\mathcal M}^-_{\mathrm{RW}}$, \citet{Bhattacharya06}
found some evidence of self-feedback only in LG: the 95\%
HPD BCRs contained zero unless $t$ increased. Further, while the
relationship of the posterior mean appeared somewhat linear in $t$,
there was no relationship  with the modeled BOLD response.
Most $\gamma_{ij}(t)$'s [with the exception of $\gamma_{13}(t)$]  were
almost invariant with  respect to time $t$, unlike the clear
oscillatory nature of the time series obtained here using ${\mathcal
  M}_{\mathrm{DP}}^{(1)}$ (or even~$\mM_{\mathrm{DP}}$). The fact that the BOLD
response had very little relationship with these effective
connectivity parameters is perplexing, given that these regions were
the ones found to be activated in the preprocessing of the fMRI
dataset. The results on $\gamma_{ij}(t)$'s using ${\mathcal M}_{\mathrm{AR}}$
were  also very surprising: while the posterior means oscillated
synchronously with $x(t)$ only for the task-irrelevant LG %and DLPFC, with
with a correlation of 0.943, % and 0.952.
there was no evidence of nonzero values for all the other effective
connectivity parameter values (including the task-relevant MOG), since
their pointwise 95\% HPD credible regions all contained zero for all
time $t$.  This is very unlike the results obtained using~%
$\mM_{\mathrm{DP}}^{(1)}$, which also established the existence of the ACN
theory in performing this task. Indeed, among all the approaches
considered in the literature and here on this dataset, only the
DP-based  analyses have been able to capture both the dynamic as well
as the oscillatory nature of the effective connectivity
parameters. In doing so, we also obtain further insight into how an
individual brain performs a Stroop task.
 %The synchronized oscillation with the BOLD response in
%these $\gamma_{ij}(t)$s is especially appealing because
%it is natural to expect and very satisfying given that it
%is picked up  despite not being specified by the
%model. There is a  slight ``drift'' in the
%$\gamma_{ij}(t)$'s matching the expectation of learning
%while performing the experiment.
%The clustering of the effective connectivity parameters is a
%scientifically important issue and our methodology, for the first
%time, sheds light on this.
%Since our model does not favour any particular clustering {\it a priori}
%the specific clustering we obtained seems to be strongly favoured
%{\it a posteriori}.
%Thus there is a clear benefit to  incorporating nonstationarity in
%the dynamic modeling of the neuronal system, for which our approach
%provides a practical and reasonable approach.
%
% The results are clearly very different from those presented in
% \ctn{Bhattacharya06} with the random walk model for
% $\gamma_{ij}(t)$. In fact Figure (5) in \ctn{Bhattacharya06}, which
% corresponds to a slightly simplified version of model $\bM_1$, shows
% that the posteriors of most $\gamma_{ij}(t)$ (with the exception of
% $\gamma_{13}(t)$) under that model may be almost invariant with
% respect to time $t$, unlike the clear oscilatory nature of the time
% series we discovered in this current work.
%The mode of the posterior distribution of $\rho$, displayed in Figure (\ref{figrhorealdatarestricted}) appears to
%be close to one, as in the case of the simulation study (see Figure (\ref{figsimstudyrhodp})).

%s5 ###
\section{Conclusions and future work}
\label{secconclusion}
Effective connectivity analysis provides an important approach to
understanding the functional organization of the human
brain. \citet{Bhattacharya06} provide a coherent and elegant Bayesian
approach to incorporating uncertainty
in the analysis. In this paper we note that this approach also brings
forth with it some limitations. In this paper we therefore propose a
nonstationary and nonparametric Bayesian approach using  a DP-based
model that embeds an AR(1) process in the class of many possible
models. Heuristically, our suggestion has some connection with  model
averaging, where we have,   a priori, an AR(1) model in mind for
specifying dynamic effective connectivity: the DP provides a coherent
way to formalize our intuition.  We have also derived an easily
implemented Gibbs sampling algorithm for learning about the posterior
distributions of all the unknown quantities. Simulation studies show
that our model is a better candidate for the analysis of effective
connectivity in many cases. The advantage is more pronounced with
increasing departures from stationarity in the true model. We also
applied our methodology to investigate the feedback mechanisms between
the task-irrelevant LG, the task-relevant MOG and the ``executive
control'' DLPFC  in the context of a~single-subject Stroop task
study. Our results showed strong self-feedback for LG and MOG, but not
for DLPFC. Further, MOG and DLPFC influence LG strongly but the
reverse is rather mild. The influence of MOG on DLPFC and vice versa
are very similar. All these discovered feedback mechanisms oscillate
strongly in the manner of the BOLD signal and are supportive of  the
framework postulated by ACN theory.
% Our results indicate, unlike what is generally believed, that the
% three different brain regions do not suppress each other in
% different ways, and that there are, in fact, 3 clusters into which
% all the 9  effective connectivity parameters fall. The clustering
% provides specific details on the nature of influences exerted by the
% brain regions on one another.
%In a nutshell, LG suppresses and is suppressed by all the three regions in a
%like manner, MOG and
%all the 3 regions inflence the region LG in the same way, in turn, LG also inflences
%the other two regions MOG and DLPFC in a like manner. The same inflence is exerted by
%MOG and DLPFC on one another. The self-feedback of the region MOG is the same as
%the inflence exerted by LG, MOG and DLPFC on LG.
%These are new findings, not reported elsewhere in the literature.
Our analysis also provides understanding into the mechanism of how the
brain performs a Stroop task. All these are novel findings not
reported in the context of fMRI analysis in the literature.
Thus, adoption of our DP-based approach not only provided interpretable
results, but---as very kindly pointed out by a reviewer---yielded additional insights into the workings of the brain.

There are several aspects of our methodology and analysis that deserve
further attention. For one, we have investigated ACN in the context of
a~Stroop task for a single male volunteer. It would be of interest to
study other tasks and responses to other stimuli and also to see how
our results on a Stroop task translate to multiple subjects and to investigate
how these mechanisms differ from one person to another. Our modeling
approach can easily be extended to incorporate such
scenarios. Further, our methodology, while developed and evaluated in
the context of modeling dynamic effective connectivity in fMRI
datasets, can be applied to other settings also,
especially in situations where the actual models for the
unknowns may be quite difficult to specify correctly. Thus, we note
that while this paper has made an interesting contribution to
analyzing dynamic effective connectivity in single-subject fMRI
datasets, several interesting questions and extensions meriting
further attention remain.

\begin{supplement}%[id=suppA]
\stitle{Contents}
\slink[doi]{10.1214/11-AOAS470SUPP}  %[doi,text={...}] - jei reikia suskaldyti doi
\slink[url]{http://lib.stat.cmu.edu/aoas/470/supplement.pdf}
\sdatatype{.pdf}
\sdescription{Section  S-1 contains additional details regarding our
  methodology, including explicit forms of the full conditional
  distributions of specific parameters, the configuration
  indicators and the distinct parameters associated with the Dirichlet
  process needed for Gibbs sampling. Detailed arguments that show
  model averaging  associated with our DP-based model   $\mM_{\mathrm{DP}}$ are
  also presented there. Section S-2 provides additional information on our
  simulation experiments, including associated methodology and results.
  Section S-3 presents further details on the analysis of the Stroop
  task experiment, including display of the data, detailed
  assessment of convergence of our  MCMC samplers when using
  $\mM_{\mathrm{DP}}$ and   $\mM^{(1)}_{\mathrm{DP}}$ and MCMC-based posterior
  analysis using $\mM_{\mathrm{AR}}$ and other additional models obtained by
  setting some effective connectivity parameters to zero. Additional
  methodological details and results regarding the smoothing of the
modeled BOLD signal $x(\cdot)$ are also presented there.}
\end{supplement}

\section*{Acknowledgments} The authors are very grateful to the Editor
and two reviewers, whose very detailed and insightful comments on
earlier versions of this manuscript greatly improved its content and
presentation.

% imsref loaded by smiklovaite, 2011-05-01 08:24:32

\printaddresses

\end{document}